\documentclass[prl,letterpaper,twocolumn,showpacs,superscriptaddress,floatfix]{revtex4}
\usepackage{graphicx,psfrag,amsmath,amssymb,amsfonts,bbm,latexsym,color,dcolumn,bm}

\begin{document}

\title{Optimized sympathetic cooling of atomic mixtures via fast adiabatic strategies}

\author{Stephen Choi}

\affiliation{Department of Physics, University of Massachusetts, Boston, MA 02125, USA}

\author{Roberto Onofrio}

\affiliation{Dipartimento di Fisica ``Galileo Galilei'', Universit\`a  di Padova, Via Marzolo 8, Padova 35131, Italy}

\affiliation{Department of Physics, University of Massachusetts, Boston, MA 02125, USA}

\affiliation{ITAMP, Harvard-Smithsonian Center for Astrophysics, 60 Garden Street, Cambridge, MA 02138, USA}

\author{Bala Sundaram}

\affiliation{Department of Physics, University of Massachusetts, Boston, MA 02125, USA}

\date{\today}

\begin{abstract}
We discuss fast frictionless cooling techniques in the framework of sympathetic cooling 
of cold atomic mixtures. It is argued that optimal cooling of an atomic species - in which 
the deepest quantum degeneracy regime is achieved - may be obtained by means of sympathetic 
cooling with another species whose trapping frequency is dynamically changed to maintain 
constancy of the Lewis-Riesenfeld adiabatic invariant. Advantages and limitations of this 
cooling strategy are discussed, with particular regard to the possibility of cooling Fermi 
gases to a deeper degenerate regime.
\end{abstract}


\maketitle
The continuous progress in achieving low temperatures has opened up new research directions 
in atomic physics, most notably the formation and study of degenerate Bose and Fermi gases. 
For the latter, most of the interesting physics, such as novel superfluid states or 
ordered phases, is conditional on the achievement of very low degeneracy factors 
$T/T_{\mathrm{F}}$, now limited to above $\simeq 10^{-2}$ \cite{Ketterle,Giorgini}. 
In most of the experiments the Fermi gas is in thermal contact with a Bose gas 
cooled with the usual evaporative cooling strategy and having a much larger heat 
capacity than the Fermi gas, resulting in the so-called sympathetic cooling 
(see \cite{Truscott,Schreck,Hadzibabic} for earlier demonstrations). 
However, both the loss of atoms intrinsic to evaporative cooling and the 
strong dependence on temperature of the heat capacity limit the efficiency 
of sympathetic cooling \cite{Truscott,Presilla}. 

If the main goal of the experiments is to achieve the deepest Fermi degeneracy, an alternative 
procedure can be envisaged with the use of species-selective traps in which the Bose gas alone 
is kept as close as possible to the nondegenerate regime, as proposed for bichromatic optical 
traps in \cite{Onofrio}, and for light-assisted magnetic traps in \cite{Brown0}. 
Recently, species-selective light-assisted magnetic traps have been successfully 
implemented to sympathetically cool Bose-Bose mixtures such as ${}^{41}$K-${}^{87}$Rb 
\cite{Catani} and ${}^{87}$Rb-${}^{174}$Yb \cite{Baumer}, and these experimental 
techniques could be naturally extended to Fermi-Bose mixtures. 
However, the ultimate, optimal ideal coolant would be a gas that retains the maximum value 
of its heat capacity, both because it does not enter the degenerate regime, and 
also because the number of atoms is not diminished during the cooling stage. 
These two features are common to a new class of cooling techniques known under 
the name of frictionless (or, less rigorously -- as seen in later considerations 
-- fast adiabatic) cooling. This method was pioneered in atomic physics in \cite{Chen}, 
and experimentally implemented for fast decompression of ${}^{87}$Rb atoms 
both in the nondegenerate \cite{Schaff0} and degenerate \cite{Schaff} regimes, 
and for demonstrating fast atomic transport \cite{Torrontegui}.

It is the purpose of this Rapid Communication to explore the extent to which frictionless cooling 
techniques may be useful in sympathetic cooling of Fermi gases, while also discussing 
their practical limitations. An important motivating factor is that an usually undesired 
feature of frictionless cooling, {\it i.e.} the fact that the atomic cloud does not 
increase its phase space density and therefore its degeneracy, is instead a crucial 
asset from the perspective of maintaining the gas in the nondegenerate regime and 
making it an ideal coolant.  

Fast frictionless cooling \cite{Chen} is based on the use of an adiabatic 
invariant for harmonically trapped atoms of mass $m$ -- hereafter assumed 
in the nondegenerate regime and therefore with negligible interactions. 
Following Lewis and Riesenfeld \cite{Lewis0,Lewis}, we consider the operator
$\hat{I}(t)=\hat{\pi}^2/2m+ m \omega_0^2 \hat{q}^2/(2 b^2)$ where 
$\hat{\pi}=b\hat{p}-m \dot{b}\hat{q}$ is the momentum operator conjugate 
to the operator $\hat{q}/b$. The parameter $\omega_0$ can be chosen as the 
initial frequency, and $b(t)$ is a time-dependent frequency scaling factor which must 
satisfy, in order for $\hat{I}(t)$ to be an invariant operator, the Ermakov equation
$\ddot{b}(t) + \omega(t)^2b = \omega_0^2/b^3$.
This can be solved by imposing boundary conditions on $b(t)$ and its first and 
second time derivatives, if we assume both a targeted final trapping frequency 
$\omega_{\mathrm f}$ and a total time duration for the adiabatic protocol $t_{\mathrm f}$ 
\cite{Chen}, thus obtaining a well-defined Ermakov trajectory for the requested time-dependent 
trapping frequency as $\omega^2(t) = \omega_0^2/b^4(t) -  \ddot{b}(t)/b(t)$.

In our analysis, we compare three decompression strategies for an harmonically trapped 
atomic cloud as shown in Fig. 1 where, for concreteness, we assume the initial trap frequency 
$\omega_0 = 2 \pi \times 250$ Hz and  the final angular frequency 
$\omega_{\mathrm f} = 2\pi  \times 2.5$ Hz, $\omega_0/\omega_{\mathrm f} = 10^2$. 
Two of the trajectories for $\omega^2(t)$ (continuous and dashed lines) are solutions of 
the Ermakov equation and therefore leave the Lewis-Riesenfeld operator invariant in time. 
The main difference in the two trajectories is the absence (continuous) or the 
presence (dashed) of a stage in which the square frequency becomes negative. 
As shown in \cite{Chen}, the presence of an {\sl anti-trapping} stage  appears 
for small targeted times $t_{\mathrm f}$. Physically, the anti-trapping stage enables the 
wave function to spread out faster so as to reach the final target width within given $t_{\mathrm f}$. 
It is interesting to note that the Ermakov equation typically produces a trajectory where, after 
anti-trapping, the trap frequency first  increases to a value above $\omega_{\mathrm f}$ before 
finally relaxing down to $\omega_{\mathrm f}$. This may be interpreted as a way to control the 
excess momentum directed away from the origin attained during the anti-trapping stage. 
Experimentally, an anti-trapping stage can be obtained by properly modulating the power of 
blue-detuned beams, as discussed in \cite{Onofrio}. For comparison, the third trajectory 
shown (dot-dashed line) is a {\sl naive} decompression strategy based on a linear ramp 
down of the frequency to its targeted final value. Its overall duration is chosen as 
$t_{\mathrm f} = T_\mathrm{max}$, where $T_\mathrm{max}= 2 \pi/\omega_{\mathrm f}$ is 
the longest harmonic oscillator period during the evolution.
In this way we compare two Ermakov-related cooling strategies to a strategy which is 
adiabatic in the usual sense of the word, {\it i.e.} with parameters changing slowly 
with respect to the intrinsic timescale set by the harmonic oscillator frequencies.  
\begin{figure}[t]
\begin{center}
\includegraphics[width=0.95\columnwidth]{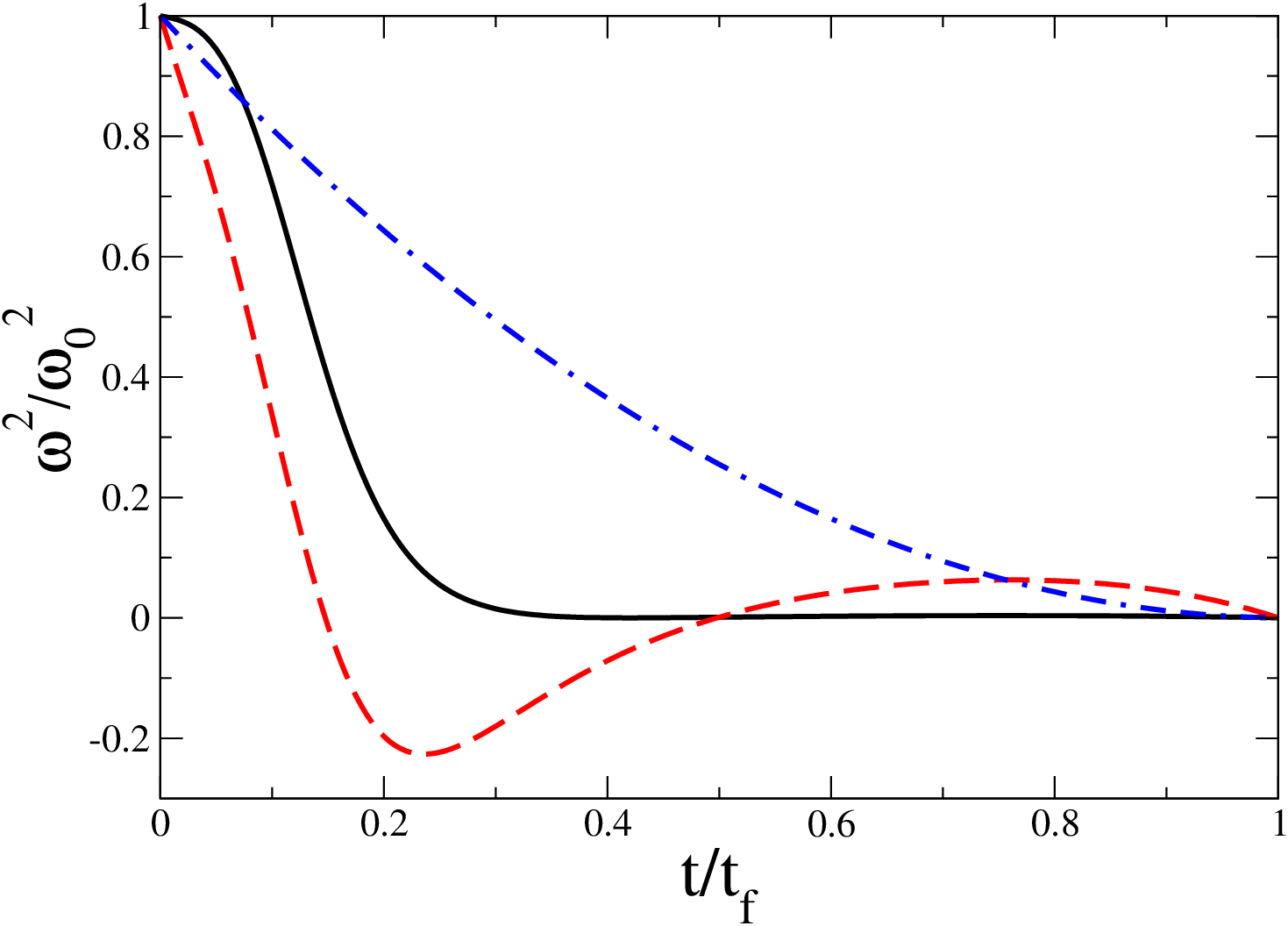}
\end{center}
\caption{(Color online) Comparison of various decompression strategies as
seen from the square of the trapping frequency, normalized to its initial 
value, versus time normalized to the overall duration of the decompression. 
The curves here and in the following figures indicate an Ermakov trajectory 
with positive square frequency lasting 25 ms (black, continuous line), a fast 
Ermakov trajectory of total duration of 6 ms (red, dashed line), and a linear ramp 
down of the trapping frequency approximating a conventional slow adiabatic strategy 
(blue, dot-dashed line) with duration of 400 ms. Notice in particular, that 
$\omega(t)^2/\omega_0^2 < 0$ in a finite interval in the 6 ms case, leading to
an anti-trapping stage, which occurs for all choices of $t_{\mathrm f}< 25$ ms.}
\label{adiabatic.fig1}
\end{figure}

We have numerically solved the Schr\"{o}dinger equation for the time-dependent 
harmonic oscillator evolving under the $\omega(t)$ solution of the Ermakov equation.  
By comparing the solution with the solution obtained from the effective Gaussian dynamics formalism 
\cite{Pattanayak} it was found that the wave function, if prepared initially as a Gaussian, retains 
its Gaussian form throughout the evolution, which proves useful in characterizing the temperature 
of the cloud. The velocity distribution is routinely used to estimate experimentally the 
temperature of ultracold atomic gases. In our case, the Gaussian state can be used to represent 
an ensemble of  atoms with a normally distributed position and momentum such that, by
assuming Maxwell-Boltzmann distribution, the variance of the velocity distribution is 
$\sigma_{v}^2 = k_{B}T/m$ where $k_B$ is the Boltzmann constant and $T$ is the 
temperature of the atomic cloud. At the same time, the use of the  variance of 
the velocity distribution, or equivalently the momentum variance, allows us to 
monitor the energy population probabilities during the cooling procedure. 
This is simply a reflection of the fact that a  larger momentum variance implies 
the occupation of a large number of {\sl excited} momentum states.  
This complements Ref. \cite{Chen} where the total energy is used as an indicator 
of temperature. The time evolution of the temperature is shown for the three cases 
in the top plot of Fig. 2. 
\begin{figure} 
\begin{center}
\includegraphics[width=0.95\columnwidth]{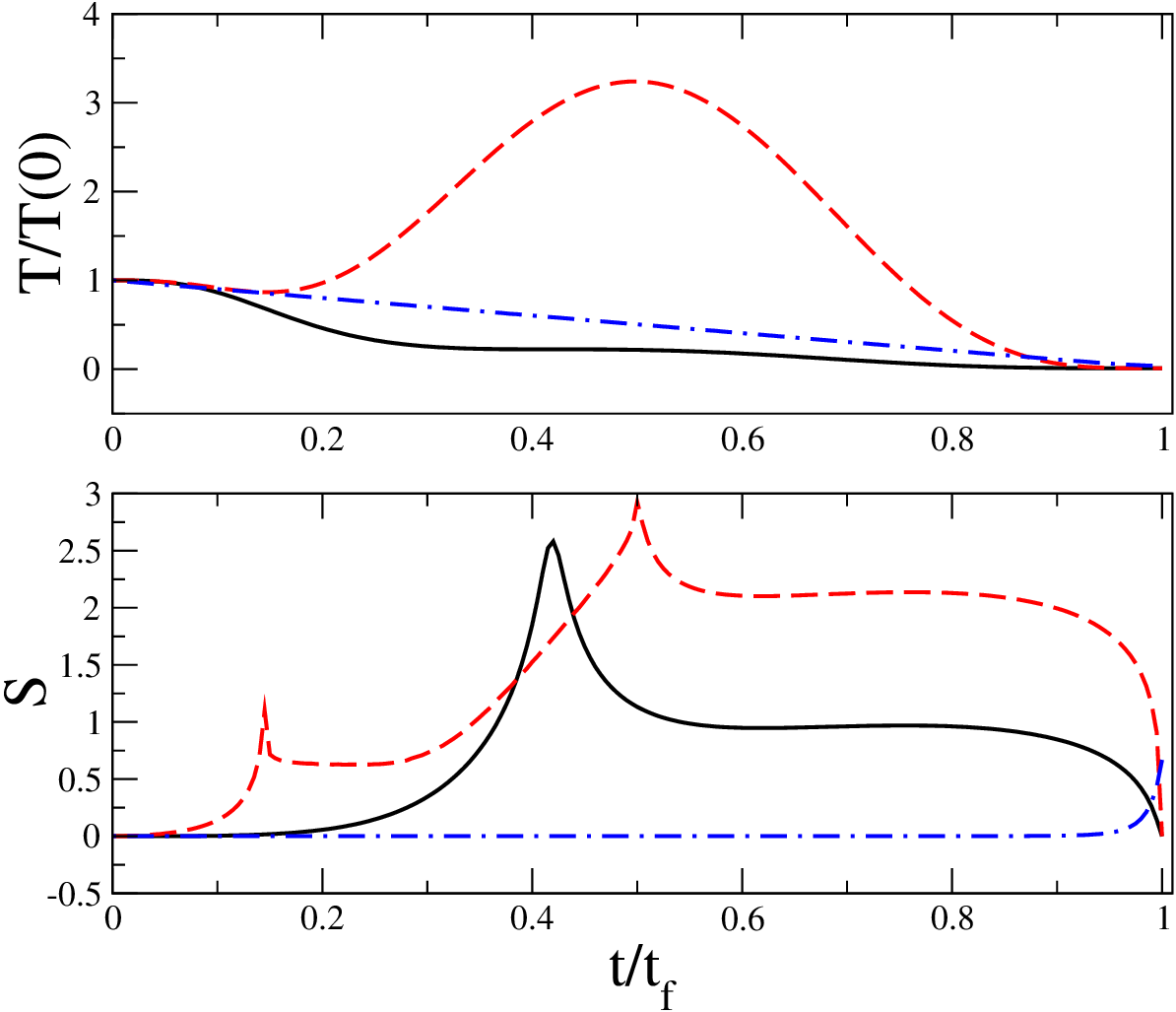}
\end{center}
\caption{(Color online) Thermodynamics of the decompression. 
Temperature scaled to initial temperature (top) and Shannon entropy 
(bottom) versus time for the three cooling strategies described in Fig. 1.
In the case of the decompression strategy corresponding to $t_{\mathrm f}= 6$ ms 
the temperature exceeds the initial temperature in a time interval 
related with the time interval in which the square frequency becomes negative, 
while in the case of $t_{\mathrm f}=25$ ms the temperature behavior is 
decreasing monotonically. The threshold situation in which the maximum temperature 
during the anti-trapping stage equals the initial temperature is achieved for 
$t_{\mathrm f} \approx 11$ ms.} 
\label{adiabatic.fig2}
\end{figure} 
In addition,  the entropy $S(t)$ is evaluated based on its Shannon definition, 
$S(t) = - \sum_{n}  |c_{n}(t)|^2 \log |c_{n}(t)|^2$ where the instantaneous 
eigenstate basis $|\psi_{n}(t) \rangle$ corresponding to the harmonic trap 
frequency $\omega(t)$ is used in the projection of the time-evolved wave function.  
Although $S(t)$ is the Shannon entropy instead of thermodynamic entropy, an adiabatic 
process would keep $c_{n}(t)$ constant in time, meaning the entropy would remain constant.

In the case of the frictionless strategy with an anti-trapping stage the basis 
for the inverted harmonic oscillator \cite{Yuce} was used ($\hbar=m=1$):
\begin{eqnarray}
\psi_{1}(x, t)&=&N\exp\left ( \frac{i\omega x^2}{2}-\frac{\omega t}{2}+\frac{i\epsilon}
{2\omega}e^{-2\omega t}\right ) \times \nonumber \\
& &  \cos(e^{-\omega t}\sqrt{\epsilon}x) \nonumber \\
\psi_{2}(x, t)&=& \exp\left ( \frac{i\omega x^2}{2}  + ik e^{-\omega t}x  -\frac{\omega t}{2}+\frac{ik^2}
{4\omega}e^{-2\omega t} \right ) \nonumber
\end{eqnarray}
where $\psi_1(x,t) $ corresponds to the solution for which the particle is confined to an expanding 
box with a moving boundary condition.  $N$ is the normalization constant  and  the discrete energy 
eigenvalues are $E_n =  e^{-2 \omega t} \epsilon$ where  $\epsilon = (n + \frac{1}{2})^2 \pi^2 /L_0^2$ 
with $n$ an integer and $L_0$ is the width of the wave function as it first enters the inverted potential.  
$\psi_2(x,t)$ corresponds to the plane wave solution with $k$ being the wave number, and since the plane 
wave solution cannot be normalized,  there is no normalization constant. It was found that, as the anti-trapping 
stage lasts for extremely short duration typically around $0.25 t_{\mathrm f}$, there is not enough time for 
our wave functions  to couple to the free-particle plane wave solution of $\psi_2(x,t)$. Therefore our wave 
functions project  entirely onto eigenstates $\psi_{1}(x,t)$.  The results of this entropy calculation are 
shown in the bottom panel of Fig. 2.  

\begin{figure}[t]
\begin{center}
\includegraphics[width=0.95\columnwidth]{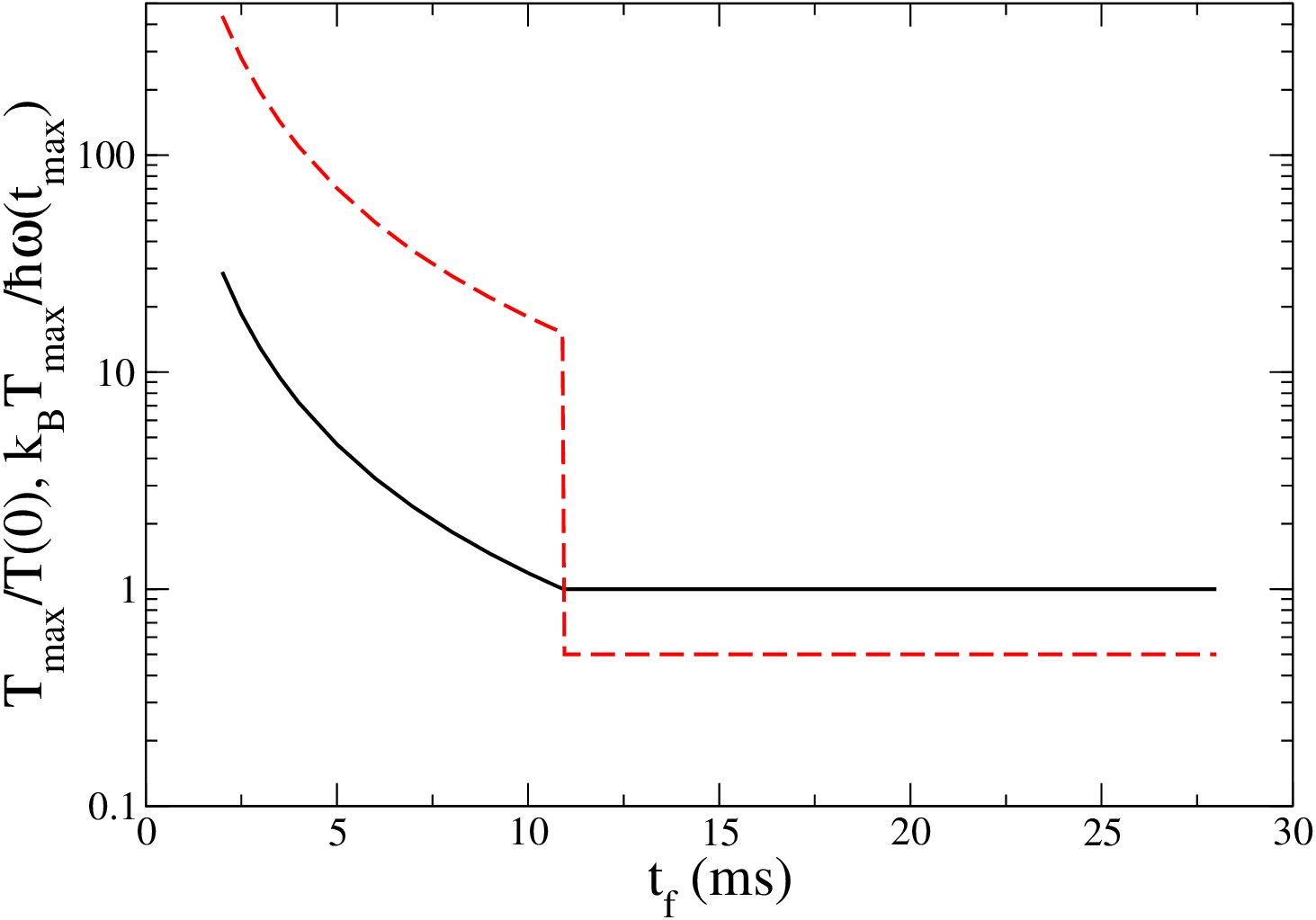}
\end{center}
\caption{(Color online) Dependence of the maximum temperature attained versus 
the duration of the cooling strategy $t_{\mathrm f}$ for the targeted temperature 
decrease $T(t_{\mathrm f})/T(0)=10^{-2}$. Solid line: maximum achieved temperature 
scaled to the initial temperature. Dashed line: maximum temperature expressed in 
units of quanta of the harmonic oscillator with the instantaneous frequency 
corresponding to the time at which the maximum temperature occurs. 
For $t_{\mathrm f}$ larger than $\approx 11$ ms, the initial temperature is also 
the maximum temperature. For shorter $t_{\mathrm f}$, substantially higher temperatures 
occur after the anti-trapping region as seen in the top plot in Fig. 2.} 
\label{adiabatic.fig3}
\end{figure}
Two general features are worth commenting on.  First, for the two cases corresponding to the 
conservation of the Lewis-Riesenfeld invariant, entropy is clearly not constant, which 
implies that adiabaticity does not hold during the cooling process. The maximum entropy 
appears at the point where the trap is shallowest while in cases where an anti-trapping 
region exists, the spikes in entropy are where the curvature of the trap changes its sign.
However, the final entropy exactly equals the initial one for these cases, as if a real 
adiabatic process has taken place. This is in stark contrast to the linear ramp down decompression 
where the entropy is constant throughout the evolution except as we approach the final  time. 
This increase is expected as the decreasing frequency leads to stricter conditions for true 
adiabaticity, indicating a need for much longer timescales for the ramp-down.

Second, the difference between the linear decrease and the Ermakov constructs is also clear in
the temperature dynamics. Higher intermediate temperatures correspond to the presence of
an anti-trapping stage, which in turn appears in the Ermakov solution for shorter $t_{\mathrm f}$.
We note that in general the presence of a stage in which the temperature of the cloud may become 
much higher than the initial cloud temperature is problematic. Since atomic traps have a finite 
potential depth, a rise of the cloud temperature will induce losses of the most energetic atoms 
\cite{Ramachandhran}. From our standpoint, this leads to a practical limitation on the minimum 
possible time for the cooling strategy, with a natural tradeoff also determined by the lifetime 
of the atoms in the trap. The final temperature for the linear ramp case is slightly higher than 
that for the other two cases. However, if the linear ramp occurred on the same time scales as the 
other two fast adiabatic methods, the final temperature would be much higher \cite{Note}.
In order to quantify this potential bottleneck of the cooling protocol, we show in Fig. 3 the 
maximum achieved temperature, {\it i.e.} the peak value of the temperature in Fig. 2, scaled 
to the initial temperature, versus the time duration of the cooling procedure. 
In the same figure we also show the maximum achieved temperature in units of quanta 
of the corresponding instantaneous harmonic oscillator $k_B T_\mathrm{max}/\hbar \omega(t_\mathrm{max})$. 
This figure of merit is useful to assess trap losses in the case of a finite trap depth. 
The results show, that for $t_{\mathrm f}$ shorter than approximately 
$11$ ms, the maximum temperature of the atoms becomes comparable to the energy of the 
15${}^\mathrm{th}$ excited state of the instantaneous harmonic trap, and quickly increases 
for shorter $t_{\mathrm f}$. Nevertheless, it appears that optimal cooling  with minimal 
trap loss on timescales comparable to the oscillation period of the trap is possible as 
long as excessively short cooling times are avoided. This is clear from the temperature curve
where for $t_{\mathrm f} > 11$ ms, the initial temperature is also the highest temperature indicating
stability in the dynamics as the trap is made more shallow. By contrast, at shorter times,
substantial heating resulting from the anti-trapping region would make the frictionless
strategy difficult to implement. The maximum heating due to anti-trapping was found to 
occur typically at  $t \simeq t_{\mathrm f}/2$,  the time when the trap changes its curvature 
at the end of anti-trapping. The sudden ``jump'' occurs at around  $t_{\mathrm f}  \approx 11$ ms, 
since for this case the maximum heating due to anti-trapping has the same magnitude as the 
initial temperature.  We emphasize that, as anti-trapping occurs for all $t_{\mathrm f}< 25$ ms 
in our example, the presence of an anti-trapping stage alone is not a sufficient condition for 
heating of the atoms to temperatures greater than the initial temperature.

A different issue arises when the spatial overlap of the two clouds is taken into account, 
as the coolant species considerably increases its size during its decompression. 
As discussed in \cite{Chen}, the position variance is directly related to the 
frequency scaling factor $b$ as $\sigma_{x}^2 = \hbar (n+1/2) b^2/(m \omega_0)$.
For the two adiabatic invariant strategies,  the temporal variation of the position variance is 
independent of $t_{\mathrm f}$  as expected from the fact that  the position variance is proportional 
to $b(t)$. Since $\sigma^{2}_x(t_{\mathrm f})/\sigma^{2}_x(0) = b^{2}(t_{\mathrm f})/b^{2}(0) = 
\omega_0/\omega_{\mathrm f} = 10^2$ as a result of the boundary condition for $b(t)$, it is 
conceivable that if a large $b(t_{\mathrm f})$ is targeted, the spread of the atomic cloud 
of the coolant will result in a small overlap with the cloud to be sympathetically cooled.  
The linear ramp down result also exhibits a large broadening expected of an adiabatic process 
that relaxes the trap frequency by a factor of $10^2$. The overlap issue seems to be crucial 
in various experiments (see for instance \cite{Hansen} for the effect of gravitational sagging 
on a large mass ratio Fermi-Bose mixture), and its effect on species-selective traps has been 
discussed in detail in \cite{Brown}. We note that, in practice, the issue of spatial overlap 
is less severe than expected, since for accidental reasons the fermionic isotope of the alkali 
primarily used in the experiments, ${}^{6}$Li, is lighter than the bosonic species used as 
coolants, ${}^{23}$Na, ${}^{87}$Rb, and bosonic isotopes of Yb. Therefore the initial overlap 
between the two species sees the fermionic species more spread out than the bosonic species. 
The decompression of the bosonic species alone in general allows for a better spatial overlap 
at intermediate times, but this also provides a limit on the maximum allowed decompression 
before the spatial overlap decreases again.  It is also worth to point out that, especially 
for very short cooling times, a possible issue arises with the sympathetic equilibration rate 
between the two species, depending on the interspecies elastic scattering rate. 
Although this issue should be studied in detail for each specific atomic mixture, in general 
this is less relevant than the spatial overlap, and can be circumvented by using magnetic or 
optical Feshbach resonances to boost the interspecies elastic scattering length as in the 
case of the  ${}^{6}$Li-${}^{87}$Rb Fermi-Bose mixture \cite{Silber,Deh,Li}.

In conclusion, we have discussed the possibility of achieving deep Fermi degeneracy via 
sympathetic cooling using a frictionless strategy for the coolant species. The identified 
advantages in this setting are the maximal heat capacity retained by the coolant due to 
the conservation of the number of atoms and the preservation of its phase space density 
in the nondegenerate regime where the specific heat retains its Dulong-Petit value. 
At the same time we have also identified two important limitations, a cooling stage 
in which the temperature rises significantly allowing for trap losses if the potential 
energy depth is not large enough, and the spreading of the cooling cloud which will reduce 
the spatial overlap with the fermionic cloud. The coolant does not need necessarily to be 
a Bose gas, as it is kept in the non-degenerate regime during the entire process, thus 
providing further flexibility in any possible experimental implementation of this scheme.


\begin{thebibliography}{99}
\bibitem{Note} We have also explored the robustness of the cooling procedure 
with respect to small deviations from the targeted frequency trajectory, by 
superimposing to it small perturbations, both of sinusoidal and broadband 
noise character. In the typical case of a sinusoidal perturbation with 
amplitude 1 $\%$ of the instantaneous frequency and period equal to 
$10^{-2} t_{\rm f}$ we obtain $T_{\mathrm f}$ differing by $0.1\%$ with 
respect to the unperturbed one.
\end{thebibliography}
\end{document}